\documentclass[prd,a4paper,preprint,preprintnumbers,nofootinbib]{revtex4-2}

\usepackage[a4paper, hdivide={1.91cm,,1.165cm}, vdivide={1.83cm,,3.0cm}]{geometry}

\usepackage{amsmath}
\usepackage{graphicx}
\usepackage{xspace}
\usepackage{color}
\usepackage{units}
\usepackage{slashed}
\usepackage[hyperfootnotes=false]{hyperref}
\hypersetup{linkcolor=red,
citecolor=green,
filecolor=cyan,
urlcolor=magenta}
\usepackage{gensymb}
\usepackage{booktabs}
\usepackage{multirow}
\usepackage{xcolor}
\usepackage{setspace}

\allowdisplaybreaks[1]

\newcommand{\ra}{\rightarrow}

\newcommand{\GG}{\langle g_s^2 G^2 \rangle}

\newcommand{\beq}{\begin{eqnarray}}
\newcommand{\eeq}{\end{eqnarray}}

\def\beq{\begin{equation}}
\def\eeq{\end{equation}}
\def\bea{\begin{eqnarray}}
\def\eea{\end{eqnarray}}

\def\vel{\left|}
\def\ver{\right|}
\def\nnb{\nonumber}
\def\lla{\left<}
\def\rra{\right>}
\def\lrar{\leftrightarrow}
\def\la{\langle}
\def\ra{\rangle}
\def\qs{\la \bar s s \ra}
\def\qu{\la \bar u u \ra}
\def\qd{\la \bar d d \ra}
\def\qq{\la \bar q q \ra}
\def\gGgG{\la g^2 G^2 \ra}
\def\es{ &=& }
\def\ar{&+& }
\def\ek{&-& }
\def\cp{&\times&}

\begin{document}


\title{Mixing angles between the tetraquark states with two heavy quarks within QCD sum rules}

\author{S.~Bilmis}
\email{sbilmis@metu.edu.tr}
\affiliation{Department of Physics, Middle East Technical University, Ankara, 06800, Turkey}
\affiliation{TUBITAK ULAKBIM, Ankara, 06510, Turkey}

\date{}

\begin{abstract}
 LHCb Collaboration has recently announced the observation of a doubly charmed tetraquark  $T_{cc \bar{u} \bar{d}}^+$ state with spin parity $J^P = 1^+$. This exotic state can be explained as a molecular state with small binding energy. According to conventional quark model, both $D^+ D^{0\ast}$  and $(D^0 D^{+\ast})$ multiquark states are expected to have the same mass and flavor in the exact $SU(3)$ symmetry. However, since the quark masses are different, $SU(3) (SU(2))$ symmetry is violated, hence, the mass and flavor eigenstates do not coincide. The mass eigenstates can be represented as a linear combination of the flavor eigenstates, which is characterized by the mixing angle $\theta$. 
In the present work,  the possible mixing angles between the $T_{cc}$ states are calculated. Moreover, the analysis are extended for all the possible tetraquarks scenarios with two heavy and two light quarks within the molecular picture although those states have not been observed yet. Our prediction on mixing angle between doubly charmed tetraquark states shows that $SU(3)$ symmetry breaking is around $7\%$ maximally.

\end{abstract}

\maketitle

\newpage


\section{Introduction}
\label{intro}
According to the Constituent Quark Model (CQM) proposed by Gell-Mann~\cite{Gell-Mann:1964ewy} and Zweig~\cite{Zweig:1964ruk} baryons are formed from three-quarks $(qqq)$ and  mesons are consisted of quark and anti-quark pairs $(\bar{q}q)$. This model has been very successful in classifying the hadrons. Many new hadrons predicted by this model have subsequently been observed with the advances of the accelerators' technologies. However, many states of the hadrons predicted by the quark model are still waiting to be discovered and form the major research area of the hadron spectroscopy. In 2003, BELLE collaboration announced the discovery of a new exotic hadron $X(3872)$ in the decay channel of $B^0 \rightarrow J/\Psi \pi^+ \pi^- K$ whose properties could not be explained by the CQM~\cite{Belle:2007hrb}. This discovery was later verified by BABAR~\cite{BaBar:2004oro}, CDF~\cite{CDF:2003cab}, D0~\cite{D0:2004zmu}, LHCb~\cite{LHCb:2011zzp} and CMS~\cite{CMS:2013fpt} collaborations. This unexpected observation increased the interest in exotic hadrons, and more than twenty exotic hadronic states have been discovered at accelerator and flavor factories till now ~\cite{Zyla:2020zbs}. These exotic states have been observed either as tetraquarks (two quarks and antiquark pairs $qq\bar{q}\bar{q}$) or as pentaquarks (four quarks and an antiquark $qqqq\bar{q}$). All the exotic hadronic states discovered up to now contained a pair of heavy valence quarks (either $\bar{c} c $ or $\bar{b}b$). No exotic state with a single heavy quark has been observed yet~\cite{Chen:2016spr}. 

The theoretical attempts for the explanation of the unexpected states can be categorized into two main approaches~\cite{Ali:2017jda,Godfrey:2008nc}. The exotic states can be tightly bound color-singlet tetraquark state ($\bar{Q}Q \bar{q}{q}$) formed by two heavy ($\bar{Q}Q$) and light ($\bar{q}q$) quark-antiquark pair states bound by a gluon. This framework is named as diquark model in the literature~\cite{Maiani:2004vq,Maiani:2004uc,Jaffe:2003sg}. Another idea is that these exotic states are weakly bound molecular states of two mesons (for tetraquarks) or a meson and a baryon (pentaquarks)~\cite{Godfrey:2008nc,Swanson:2006st,Karliner:2015ina}. The mass and decay widths of the exotic hadrons can be calculated in diquark and molecular state models, which needs to be confirmed with the experiments. The properties of the exotic states both from the theoretical and experimental perspectives are discussed widely in the literature (see reviews ~\cite{Guo:2017jvc,Agaev:2020zad,Godfrey:2008nc,Lebed:2016hpi,Esposito:2016noz}). 

One of these exotic hadrons, namely, very narrow $T_{cc}^+$ tetraquark state in the $D^0 D^0 \pi^+$ spectrum recently has been observed by LHCb Collaboration~\cite{LHCb:2021vvq,LHCb:2021auc}. This is the first experimental evidence of the open double charmed tetraquark with $cc\bar{u}\bar{d}$ quark configuration. The spin-parity of $T_{cc}^+$ state is determined as $J^P = 1^+$ and the measured mass of the tetraquark $T_{cc}^+$ is located at $(-273 \pm 61 \pm 5_{-14}^{+11})~keV$ just below the $D^0 D^{+ \ast}$ mass threshold. For this reason, the molecular picture is quite attractive for studying the properties of the $T_{cc}^+$ state~\cite{Li:2012ss,Xu:2017tsr,Li_2021,Ren:2021dsi,Chen:2021vhg,Wu:2021kbu,Chen:2021tnn,Wang:2020rcx}.

The interactions between $D^0 D^{*+}$ and  $D^+ D^{*0}$ are practically the same. Hence, if  $T_{cc}^+$ were described as the $D^0 D^{*+}$ molecule, there should also exist the other partner  molecule $D^+ D^{*0}$. It is a well-known fact that the mixing takes place if two states have the same total angular momentum and parity, i.e., $J^P$. Since the states, $D^0 D^{*+}$ and $D^+ D^{*0}$ carry the same $J^P$, mixing between these two states is expected. A similar argument can also be made about the existence of mixing angles between the $QQ\bar{q}_1 \bar{q}_2$ states, where $Q$ is the heavy $c(b)$ quark, $q_1$ and $q_2$ are the light $u,~d$ and $s$ quarks by anticipating the existence of $T_{bb}$ states.

In the present work, we calculate the mixing angles between  
$(\bar{q}_1 \gamma_5 Q) (\bar{q}_2 \gamma_\mu Q)$ and 
$(\bar{q}_2 \gamma_5 Q)  (\bar{q}_1 \gamma_\mu Q)$ states within the QCD
sum rules method by following the approach introduced in \cite{Aliev:2010ra}, assuming that these states are molecular states. Possible measurement of the mixing angle can indirectly mimic the nature of the tetraquark state as a hadronic molecule.

The paper is organized as follows. In Section \ref{sec:2}, the theoretical calculations on the mixing angle between tetraquark states are performed. The Section~\ref{sec:3} is devoted to the numerical analysis of this quantity, and the last section contains our discussion and conclusion.
\section{Determination of the mixing angles between the $(\bar{q}_1 \gamma_5 Q)
(\bar{q}_2 \gamma_\mu Q)$ and $(\bar{q}_2 \gamma_5 Q)
(\bar{q}_1 \gamma_\mu Q)$ states}
\label{sec:2}

In determination of the mixing angles between the 
$(\bar{q}_1 \gamma_5 Q) (\bar{q}_2 \gamma_\mu Q)$ and
$(\bar{q}_2 \gamma_5 Q)  (\bar{q}_1 \gamma_\mu Q)$ states
in the framework of the QCD sum rules method, we start by considering the
following correlation function,
\bea
\label{ema01}
\Pi_{\mu\nu} = i \int d^4x e^{ipx} \lla 0 \vel T\left\{ J_{\mu}^{(1)} (x) \bar{J}_{\nu}^{(2)} (0)
\right\} \ver 0 \rra~.
\eea
This correlation function can be written in terms of two independent invariant functions as,
\begin{equation}
  \label{eq:a1}
  \Pi_{\mu \nu}(p^2) = (g_{\mu \nu} - \frac{p_\mu p_\nu}{p^2}) \Pi_1 (p^2) + \frac{p_\mu p_\nu}{p^2} \Pi_2(p^2)
\end{equation}
where the first and second structures describe the contribution of spin-1 and spin-0 states, respectively. Here $J_{1\mu}$ and $J_{2\nu}$ are the interpolating currents of the corresponding physical states that can be written as linear combinations of unmixed states as
\bea
\label{ema02}
J_{\mu}^{(1)} (x) \es \cos\theta j_\mu^{(1)} + \sin\theta  j_\mu^{(2)}~, \nnb \\
J_{\nu}^{(2)} (x) \es - \sin\theta j_\nu^{(1)} + \cos\theta  j_\nu^{(2)}~,
\eea
where
\bea
j_{\mu}^{(1)} (x) \es (\bar{q}_1^a \gamma_5 Q^a) (\bar{q}_2^b \gamma_\mu Q^b)~,\nnb \\
j_{\nu}^{(2)} (x) \es (\bar{q}_2^a \gamma_5 Q^a) (\bar{q}_1^b \gamma_\nu Q^b)~. \nnb
\eea
correspond to unmixed states. Here $a$ and $b$ are the color indices.

Now let us introduce the correlation function corresponding to the unmixed states, i.e,
\begin{equation}
  \label{eq:a2}
  \tilde{\Pi}_{\mu \nu}^{i j} = i \int d^4 x e^{i p x } \langle 0 | j_\mu^{i} j_\nu^{j} |0 \rangle
\end{equation}
where $i$ and $j$ runs from 1 to 2. Again separating the contribution of spin-0 and spin-1 states, this correlation function can be written as
\begin{equation}
  \label{eq:a3}
  \tilde{\Pi}^{i j}_{\mu \nu}(p^2) = (g_{\mu \nu} - \frac{p_\mu p_\nu}{p^2}) \tilde{\Pi}_1^{i j} (p^2) + \frac{p_\mu p_\nu}{p^2} \tilde{\Pi}_2^{i j}(p^2)
\end{equation}
In the following discussions, we will only consider the coefficient of the structure $(g_{\mu \nu} - \frac{p_\mu p_\nu}{p^2})$ since all the considered states are assumed to have quantum numbers $J^{P} = 1^+$. 

The sum rules for the quantity under consideration can be obtained 
by calculating the correlation function in two
different regions, i.e., in terms of hadrons and in terms of
quarks--gluons in the deep Euclidean domain, and matching the results 
of the two representations of the correlation function.
The phenomenological part of the correlation function can be
obtained by saturating it with hadron states carrying the same quantum
numbers as the interpolating current and then isolating the ground state
contributions. The currents $J_{1\mu}$ and $J_{2\mu}$ are created from the
vacuum states of the corresponding mesons, respectively, and hence the
phenomenological part of the correlation function should be equal to zero.
In other words, the mixing angle is solely determined in terms of the quark
and gluon degrees of freedom. As a result, the mixing angle is free from the
uncertainties coming from the hadronic part.

We now turn our attention to the calculation of the theoretical part of the
correlation function. 

Using Eqs.(\ref{ema01}) and (\ref{ema02}), we get,
\bea
\label{ema03}
-\sin\theta \cos\theta \,\widetilde{\Pi}_{\mu\nu}^{(11)} + \cos^2\!\theta \,\widetilde{\Pi}_{\mu\nu}^{(12)}
-\sin^2\!\theta \, \widetilde{\Pi}_{\mu\nu}^{(21)} + \sin\theta \cos\theta \, \widetilde{\Pi}_{\mu\nu}^{(22)}
= 0~.
\eea
Choosing the coefficient of the structure  $(g_{\mu \nu} - \frac{p_\mu p_\nu}{p^2})$, which only contains the contribution
of $J^P = 1^+$ state, we get from Eq.(\ref{ema03}),
\bea
\sin\theta \cos\theta \left (\widetilde{\Pi}^{(22)} - \widetilde{\Pi}^{(11)}
\right) + \cos^2\!\theta \,\widetilde{\Pi}^{(12)}
- \sin^2\!\theta \,\widetilde{\Pi}^{(21)} = 0~. \nnb
\eea
Dividing both sides to $\cos^2\!\theta$ (assuming that $\cos\theta \neq 0$),
and solving the quadratic equation for $\tan\theta$ we get,
\bea
\label{ema04}   
\tan\theta = { \widetilde{\Pi}^{(22)} - \widetilde{\Pi}^{(11)} \pm
\sqrt{ \left( \widetilde{\Pi}^{(22)} - \widetilde{\Pi}^{(11)} \right)^2 +
4 \widetilde{\Pi}^{(12)} \widetilde{\Pi}^{(21)} }  \over
2 \widetilde{\Pi}^{(21)} }~.
\eea
We already noted that for obtaining the sum rules for the relevant quantity, the correlation function (Eq.\ref{ema01}) should be calculated in terms of quarks and
gluons in the deep Euclidean region $p^2 \ll 0$ by using the operator
product expansion (OPE). Its expression can be obtained by substituting Eq.(\ref{ema02}) into Eq.(\ref{ema01}) and then using the Wick theorem. As a result, we obtain the correlation function in terms of the light
and heavy quark propagators and the light quark condensates. So, to express the correlation function in terms of the quark and gluon degrees of freedom, we need the expressions of the heavy and light quark
propagators.   

The light-quark propagators, to first order in the light quark mass, is
calculated in \cite{Ioffe:1983ju,Chiu:1986cf}, whose expression is given as
\bea
\label{ema06}
S_q^{ab}(x) \es {i \rlap/x\over 2\pi^2 x^4} \delta^{ab} - 
{m_q\over 4 \pi^2 x^2} \delta^{ab} -
{\qq \over 12} \left(1 - i {m_q\over 4} \rlap/x \right) \delta^{ab} -
{x^2\over 192} m_0^2 \qq   \left( 1 -
i {m_q\over 6}\rlap/x \right) \delta^{ab}\nnb \\
\ar{i\over 32 \pi^2 x^2} g_s G_{\mu\nu}^{ab} (\sigma^{\mu\nu} \rlap/x + \rlap/x
\sigma^{\mu\nu}) - {4 \pi \over 3^9\, 2^{10}}  \qq \GG x^2 \delta^{ab} + \cdots
\eea
The heavy-quark propagator in x--representation is given as \cite{Huang:2012ti}, 
\bea
\label{ema07}
S^{ab}_Q(x) \es {m_Q^2 \delta^{ab} \over (2 \pi)^2} \left[
i \rlap/x { K_2(m_Q\sqrt{-x^2}) \over \ (\sqrt{-x^2})^2 } +
{ K_1(m_Q\sqrt{-x^2}) \over \sqrt{-x^2} } \right] \nnb \\
\ek {m_Q  g_s G_{\mu\nu}^{ab} \over 8 (2\pi)^2}
\left[ i (\sigma^{\mu\nu} \rlap/x + \rlap/x
\sigma^{\mu\nu}) { K_1(m_Q\sqrt{-x^2}) \over \sqrt{-x^2} }
+ 2 \sigma^{\mu\nu} K_0(m_Q\sqrt{-x^2}) \right] \nnb \\
\ek  {\GG \delta^{ab} \over (3^2\,2^8 \pi)^2} \left[
(i m_Q \rlap/x - 6) { K_1(m_Q\sqrt{-x^2}) \over \sqrt{-x^2} }
+ m_Q x^4 { K_2(m_Q\sqrt{-x^2}) \over (\sqrt{-x^2})^2 } \right]~,
\eea
where $G^{\mu\nu}$ is the gluon field strength tensor, $g_s$ is the strong coupling constant and $K_0$, $K_1$ and 
$K_2$ are the modified Bessel functions of the second kind.

The invariant functions $\widetilde{\Pi}^{(ij)}$ can be related to their imaginary part (spectral density)
with the help of the dispersion relation,
\bea
\label{ema08}
\widetilde{\Pi}^{(ij)} = \int_{s_{min}}^\infty {\rho^{(ij)}(s) \over s-p^2} ds~,
\eea
where $s_{min} = (2 m_Q + m_{q_1} + m_{q_2})^2$. The spectral densities $\rho^{(12)}(s)$ and $\rho^{(21)}(s)$ are calculated in this work and their explicit expressions are presented in Appendix~\ref{appendix:a}. The spectral densities $\rho^{(11)}(s)$ and $\rho^{(22)}(s)$ are already calculated in \cite{Aliev:2021dgx}. 
Performing the Borel transformation over the variable $-p^2$, and assuming
the quark hadron duality, we get from Eq.(\ref{ema08}),
\bea
\label{ema09}
\widetilde{\Pi}^{(ij)(B)} =  \int_{s_{min}}^{s_0} \rho^{(ij)}(s) e^{-s/M^2} ds~,
\eea
where $M^2$ is the Borel mass parameter and $s_0$ is the continuum
threshold. Substituting Eq.(\ref{ema09}) into Eq.(\ref{ema04}) we obtain the
expression of the mixing angle in terms of the quark and gluon degrees of
freedom.

\section{Numerical Analysis}
\label{sec:3}
Having obtained the expression for the mixing angle, we are ready to perform
the numerical analysis in the framework of the QCD sum rules. For this purpose,
we need the values of some input parameters, which are presented 
in Table~\ref{tab:1}. For the heavy quark-masses, $\overline{\mbox{MS}}$ values are used.

\begin{table*}[h]
  \centering
  \renewcommand{\arraystretch}{1.4}
  \setlength{\tabcolsep}{7pt}
  \begin{tabular}{lr}
    \toprule
     Parameters             & Value         \\
    \midrule
    $\overline{m}_u(2~GeV)$          & $ (2.2_{-0.4}^{+0.6})~MeV$         ~\cite{PhysRevD.98.030001} \\
    $\overline{m}_d(2~GeV)$          & $ §(4.7_{-0.4}^{+0.8})~MeV$        ~\cite{PhysRevD.98.030001} \\
    $\overline{m}_s(1~GeV)$          & $(0.114 \pm 0.021)~GeV$            ~\cite{PhysRevD.98.030001} \\ 
    $\overline{m}_c(\overline{m}_c)$ & $(1.28 \pm 0.03)~GeV$              ~\cite{PhysRevD.98.030001} \\
    $\overline{m}_b(\overline{m}_b)$ & $(4.18 \pm 0.03)~GeV$              ~\cite{PhysRevD.98.030001} \\ 
    $\qu (1~GeV)$                    & $(-246_{-19}^{+28}~MeV)^3$         ~\cite{Gelhausen:2014jea} \\
    $\qd (1~GeV)$                    & $(1+\gamma) \qu~GeV^3$             ~\cite{Ioffe:1981kw} \\
    $m_0^2$                          & $(0.8 \pm 0.1)~GeV^2$              ~\cite{Ioffe:2002ee} \\
    $\GG$                            & $ 4 \pi^2 (0.012 \pm 0.006)~GeV^4$ ~\cite{Ioffe:2002ee} \\
    $\qs$                            & $(0.8 \pm 0.2) \qu$                ~\cite{Ioffe:2002ee} \\
    $\gamma$                         & $ -0.003 \div 0.01$                ~\cite{Jin:1994jz} \\
    \bottomrule
  \end{tabular}
  \caption{The values of the input parameters used in our calculations.}
  \label{tab:1}
\end{table*}
%

The sum rules contain two auxiliary parameters, namely, Borel mass square $M^2$, and the continuum threshold $s_0$. Therefore the so-called working regions of these two parameters must be determined in a way that the mixing angle exhibits good stability with respect to the variation of these parameters, respectively.

The lower and upper bounds of the Borel mass parameter $M^2$ are determined by requiring that the OPE should be convergent and pole contribution is dominant with respect to the continuum one. In other words, the upper bound of $M^2$ is obtained from the condition that the pole contribution should be  more than $50 \%$, i.e,
\begin{equation}
  \label{eq:pole}
  \text{pole contribution} = \frac{\int_{s_{\text{min}}}^{s_0} \rho(s) e^{-s/M^2} ds }{\int_{s_{\text{min}}}^{\infty} \rho(s) e^{-s/M^2} ds} > 0.5~. 
\end{equation}

To obtain the lower bound for $M^2$, we restrict the total condensate contributions to be less than $30 \%$ of the result, i.e.,
\begin{equation}
  \label{eq:lower}
  \frac{\sum \Pi_i^{\text{condensates}}}{\Pi_{\text{total}}} < 30 \%
\end{equation}

These condition lead us to the working regions of $M^2$ that are presented in Table~\ref{tab:2}.

On the other hand, the threshold value $s_0$ is determined by requiring that the variation in the obtained mass value of the considered hadron should be minimum. Using the working region of $M^2$, we find that mass sum rules exhibits very good stability on variation of the $s_0$ that are presented in Table~\ref{tab:2}.

\begin{table*}[h]
\centering
\renewcommand{\arraystretch}{1.4}
\setlength{\tabcolsep}{7pt}
\begin{tabular}{lcc}
    \toprule
                                & $s_0~(GeV^2)$     & $M^2~(GeV^2)$  \\
\midrule
$D^+ D^{0\ast}   (D^0 D^{+\ast})$   &  $20 \div 21$   &  $2.4 \div  2.9$    \\  
$D^+ D_s^{+\ast} (D_s^+ D^{+\ast})$ &  $20 \div 21$   &  $2.4 \div  2.9$    \\ 
$D^0 D_s^{+\ast} (D_s^+ D^{0\ast})$ &  $20 \div 21$   &  $2.4 \div  2.8$    \\ 
$B^0 B^{-\ast}   (B^- B^{0\ast})$   & $115 \div 120$  & $8 \div 11$    \\ 
$B^0 B_s^{0\ast} (B_s^0 B^{0\ast})$ & $120 \div 125$  & $8 \div 12$    \\
$B^- B_s^{0\ast} (B_s^0 B^{-\ast})$ & $120 \div 125$  & $8 \div 12$    \\
\bottomrule
\end{tabular}
\caption{The working regions the Borel mass parameter $M^2$, and the
continuum threshold $s_0$ for different tetraquark states in molecular
picture} \label{tab:2}
\end{table*}

\begin{figure}[h]
  \centering
\includegraphics[width=0.7\textwidth]{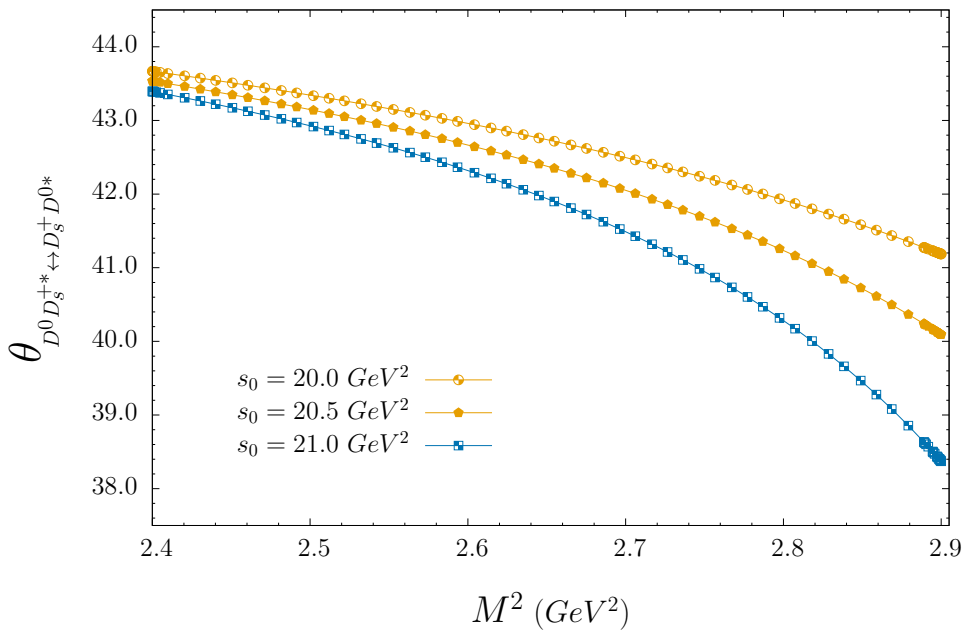}
\caption{The dependency of the mixing angles between $D^0 D_s^{+*}$ and $D_s^+ D^{0*}$ on Borel mass square at the fixed values of $s_0$.}
\label{fig:1}
\end{figure}

Having the values of input parameters and working regions of $M^2$ and $s_0$, we can perform numerical analysis for the mixing angles. In Figure~\ref{fig:1}, we present the dependency of the mixing angle between $D^0 D_s^{+*}$ and $D_s^+ D^{0*}$ on Borel mass square at the fixed values of $s_0$. Similar analyses are performed for all other mixing angles, and the results are collected in Table~\ref{tab:3}.
\begin{table*}[h]
\centering
\renewcommand{\arraystretch}{1.4}
\setlength{\tabcolsep}{7pt}
\begin{tabular}{lcc}
\toprule
                                & $\Delta \theta^{\degree} = |\theta^{\degree} - 45^{\degree}| $      \\
\midrule
$\Delta \theta_{ D^+ D^{0\ast}   \lrar D^0 D^{+\ast} }  $  &  $0.20 \pm 0.05$   \\
$\Delta \theta_{ D^+ D_s^{+\ast} \lrar D_s^+ D^{+\ast} }$  &  $2.8 \pm 0.8$   \\
$\Delta \theta_{ D^0 D_s^{+\ast} \lrar D_s^+ D^{0\ast} }$  &  $3.1 \pm 0.9$   \\
$\Delta \theta_{ B^0 B^{-\ast}   \lrar B^- B^{0\ast} }  $  &  $0.019 \pm 0.006$   \\
$\Delta \theta_{ B^0 B_s^{0\ast} \lrar B_s^0 B^{0\ast} }$  &  $0.36 \pm 0.03$    \\
$\Delta \theta_{ B^- B_s^{0\ast} \lrar B_s^0 B^{-\ast} }$  &  $0.37 \pm 0.03$   \\

\bottomrule
\end{tabular}
\caption{The values of the mixing angles between possible tetraquark states.}
\label{tab:3}
\end{table*}

The mixing angle between $D^+ D^{0\ast}$ and $D^0 D^{+\ast}$ system has been estimated within one boson exchange framework \cite{Chen:2021vhg}. However, the obtained value $\theta = \pm 30.8^{\degree}$ is considerably smaller than our result.
It should be noted that the deviation from $\tan\theta = \pm 1$ is due to the
isospin symmetry breaking, and in our case, this violation is small for the
$D^+ D^{\ast 0}$ and  $D^0 D^{\ast +}$ systems. Our results show that
the mixing angles that deviate relatively considerable from $\theta= \pm 45^0$ are only for the 
$D_s D^{\ast }$ and  $D D_s^{\ast}$ tetraquark systems.

\section{Conclusion}
\label{sec:conclusion}
Recently, LHCb Collaboration announced the observation of a new type of hadronic state, $T_{cc}$, containing two charmed and anti-$u$ and anti$-d$ quarks in the $D^0 D^0 \pi^+$ mass spectrum slightly below the $D^0D^{*+}$ threshold with quantum numbers $J^P = 1^+$. Analysis conducted in~\cite{Wang:2020rcx,Li:2012ss,Xu:2017tsr} show that the molecular picture can successfully describe this exotic state. Moreover, the quark model predicts the existence of similar states with two heavy quarks and the same quantum numbers. It is a well-known fact that the states having the same quantum numbers in principle can be mixed. Inspired by this fact, the mixing angles between tetraquark systems with two heavy quarks in the molecular picture are calculated in the framework of the QCD sum rules method. Inspired by the discovery of $T_{cc}$ state, we also studied the mixing angles for B-meson molecules for the possible $T_{bb}$ state that has not been observed yet. Our predictions on the mixing angles show that the violation of isospin symmetry leads to a very small deviation from $45^{\degree}$, which corresponds to the exact isospin symmetry. The deviation from $45^{\degree}$ is relatively large  especially for $D_s D^{\ast}$ and  $D D_s^{\ast}$ systems.  Hopefully, these findings will be tested in future LHCb experiments as well as flavor factories and provide useful information for understanding the inner structures of the tetraquark systems with two heavy quarks.     
\bibliographystyle{utcaps_mod}
\bibliography{mybib.bib}

\appendix
\section{The expression of the spectral densities}
\label{appendix:a}
The spectral densities $\rho_{12}(s)$ that are calculated in this work are shown in this appendix.  The expressions of the $\rho_{11}(s)$ can be found in \cite{Aliev:2021dgx}. Note that the spectral density $\rho_{21}(s)$ can be obtained from $\rho_{12}(s)$ by means of the replacement $( d \lrar u )$, respectively.

\bea
\rho_{12}^{(pert)}(s) \es
   {1\over 2^{14} \pi^6}
\int_{\alpha_{min}}^{\alpha_{max}} {d\alpha \over \alpha^3}
 \int_{\beta_{min}}^{\beta_{max}} {d\beta \over \beta^3}
 \Big[ (\alpha + \beta) m_b^2 - \alpha \beta s \Big]^2 \nnb \\
\cp  \Bigg\{ (1 - \beta) \beta m_b^3 \Big( \big[4 - \beta (5 + \beta)
\big] m_b + 4 \beta (3 + \beta) (m_d + m_u) \Big) \nnb \\
\ar  \alpha^4 (m_b^2 - \beta s)^2 + 2 \alpha m_b
    \Big[ m_b \Big( \Big\{ 2 - \beta \big[9 - 2 \beta (3 + \beta) \big] \Big\} m_b^2 \nnb \\
\ek 4 \beta \big[3 - \beta (3 + \beta) \big] m_d
        m_u + 2 \beta \big[3 - \beta (4 + 3 \beta) \big] m_b (m_d + m_u) \Big) \nnb \\
\ek     (1 - \beta) \beta \Big\{ \big[ 2 - \beta (3 + \beta) \big] m_b + 2 \beta (3 + \beta) (m_d + m_u)
\Big\}  s\Big] \nnb \\ 
\ar   2 \alpha^3 (m_b^2 - \beta s) \big[2 (1 + \beta) m_b^2 - 2 \beta m_b (m_d + m_u) + 
\beta (4 m_d m_u - \beta s) \big] \nnb \\
\ek  \alpha^2 \Big[3 \big\{3 - 2 \beta (2 + \beta)\big\} m_b^4 +
     4 \beta (2 + 3 \beta) m_b^3 (m_d + m_u) \nnb \\
\ek 8 \beta^2 (1 + \beta) m_b (m_d + m_u) s +
     \beta^2 s \big[ 8 \beta m_d m_u +
 (1 - \beta^2) s \big] \nnb \\
\ar 2 \beta m_b^2 \Big( 4 (3 + 2 \beta) m_d m_u + 
\big[ 5 - \beta (4 + 3 \beta)\big] s\Big) \Big] \Bigg\}
\nnb \\ \nnb \\
%
\rho_{12}^{(\qd \qu)}(s) \es {\qd \qu \over (3 \times 2^8) \pi^2}
\int_{\alpha_{min}}^{\alpha_{max}} d\alpha
   \Bigg\{ 6 m_b (m_d + m_u) + \alpha \Big[ 6 m_0^2 + 8 m_b^2 - 3 m_d m_u \nnb \\
\ek 6 m_b (m_d + m_u) - 8 s \Big] - 
\alpha^2 (6 m_0^2 - 3 m_d m_u - 8 s) \Bigg\} \nnb \\ \nnb \\
%
\rho_{12}^{(m_0^2 \qd \gGgG)}(s) \es
- { m_0^2 \qd \gGgG  \over 2^{13} m_b \pi^4}
\int_{\alpha_{min}}^{\alpha_{max}} d\alpha
(1 - \alpha)^2 \nnb \\
%
\rho_{12}^{(m_0^2 \qu \gGgG)}(s) \es
- { m_0^2 \qu \gGgG \over  2^{13}  m_b \pi^4}
\int_{\alpha_{min}}^{\alpha_{max}} d\alpha
(1 - \alpha)^2 \nnb \\
%
\rho_{12}^{(m_0^2 \qd)}(s) \es
  {m_0^2 \qd  \over (3 \times 2^{10}) \pi^4}
\int_{\alpha_{min}}^{\alpha_{max}} {d\alpha \over \alpha}
         \Bigg\{ m_b \Big[ 6 m_b^2 - 2 \alpha^2 m_b (m_d - 3 m_u) +
     3 (1 - \alpha) \alpha m_d m_u\Big] \nnb \\
\ek 2 (1 - \alpha) \alpha (3 m_b - \alpha m_d + 3 \alpha m_u) s \Bigg\} \nnb \\
\ek {m_b m_0^2 \qd \over 2^{10} \pi^4}
\int_{\alpha_{min}}^{\alpha_{max}} {d\alpha \over \alpha} \int_{\beta_{min}}^{\beta_{max}} d\beta
\Big[(\alpha + \beta) m_b^2 - \alpha \beta s \Big] \nnb \\ \nnb \\
%
\rho_{12}^{(m_0^2 \qu)}(s) \es
 { m_0^2 \qu \over (3 \times 2^{10}) \pi^4}
\int_{\alpha_{min}}^{\alpha_{max}} {d\alpha \over \alpha}
  \Bigg\{ m_b \Big[ 6 m_b (m_b + \alpha^2 m_d) -
     \alpha (2 \alpha m_b - 3 m_d + 3 \alpha m_d) m_u \Big] \nnb \\
\ek 2 (1 - \alpha) \alpha (3 m_b + 3 \alpha m_d - \alpha m_u) s \Bigg\} \nnb \\
\ek { m_b m_0^2 \qu \over 2^{10} \pi^4}
\int_{\alpha_{min}}^{\alpha_{max}}
{d\alpha \over \alpha} \int_{\beta_{min}}^{\beta_{max}} d\beta
\Big[ (\alpha + \beta) m_b^2 - \alpha \beta s \Big] \nnb \\ \nnb \\
%
\rho_{12}^{(\qd \gGgG)}(s) \es
 - {\qd \gGgG \over (3^2\times 2^{13}) m_b \pi^4}
\int_{\alpha_{min}}^{\alpha_{max}} {d\alpha \over \alpha}
\Bigg\{ 2 \big[54 - 5 \alpha (1 + \alpha)\big] m_b^2 \nnb \\
\ek 3 \alpha m_b \Big( (38 - 31 \alpha) m_d - 2 \big[ 11 - (5 - \alpha) \alpha \big] m_u \Big) \nnb \\
\ar 2 (1 - \alpha) \alpha \big[ 9 (1 - \alpha) m_d m_u - 2 (6 - 5 \alpha) s \big] \Bigg\} \nnb \\
\ek {\qd \gGgG \over (3^2 \times2^{12}) m_b \pi^4}
\int_{\alpha_{min}}^{\alpha_{max}} {d\alpha \over \alpha}
\int_{\beta_{min}}^{\beta_{max}} d\beta
 \Big[ (60 + 4 \alpha + 6 \beta) m_b^2 - 3 \alpha m_b (m_d - 2 m_u) +
    6 \alpha \beta s \Big] \nnb \\ \nnb \\
%
\rho_{12}^{(\qu \gGgG)}(s) \es
 - {\qu \gGgG \over (3^2 \times 2^{13})  m_b \pi^4}
\int_{\alpha_{min}}^{\alpha_{max}} {d\alpha \over \alpha}
   \Bigg\{2 \big[54 - 5 \alpha (1 + \alpha) \big] m_b^2 \nnb \\
\ar 3 \alpha m_b \Big( 2 \big[ 11 - (5 - \alpha) \alpha \big] m_d -
(38 - 31 \alpha) m_u \Big) \nnb \\
\ar 2 (1 - \alpha) \alpha \big[ 9 (1 - \alpha) m_d m_u -
2 (6 - 5 \alpha) s \big] \Bigg\} \nnb \\
\ek {\qu \gGgG \over (3 \times 2^{12}) m_b \pi^4}
\int_{\alpha_{min}}^{\alpha_{max}} {d\alpha \over \alpha}
\int_{\beta_{min}}^{\beta_{max}} d\beta
\Big\{m_b \Big[ (60 + 4 \alpha + 6 \beta) m_b + 6 \alpha m_d - 3 \alpha m_u
\Big] +
    6 \alpha \beta s \Big\} \nnb \\ \nnb \\
%
\rho_{12}^{(\gGgG^2)}(s) \es
 - {\gGgG^2 \over (3^3 \times 2^{18}) m_b^2 \pi^6}
\int_{\alpha_{min}}^{\alpha_{max}} {d\alpha \over \alpha}
     \Bigg\{ \big[ 192 + \alpha (293 - 116 \alpha) \big] m_b^2 \nnb \\
\ek 12 \alpha (7 + 16 \alpha) m_b (m_d + m_u) -
   12 \Big( \alpha \big[ 1 - 2 (3 - \alpha) \alpha \big] m_d m_u \nnb \\
\ar (2 - \alpha) (1 - \alpha) \alpha s \Big) \Bigg\} \nnb \\
\ar {\gGgG^2 \over (3^3 \times 2^{17}) m_b^2 \pi^6}
\int_{\alpha_{min}}^{\alpha_{max}} {d\alpha \over \alpha}
 \int_{\beta_{min}}^{\beta_{max}} {d\beta \over \beta}
    \Bigg\{ 6 \beta (33 + 16 \beta) m_b^2 +
   \alpha m_b \big[ (18 + 53 \beta) m_b \nnb \\
\ar 12 \beta (m_d + m_u) \big] - 18 \alpha \beta s \Bigg\} \nnb \\ \nnb \\
\rho_{12}^{(\gGgG)}(s) \es
 {\gGgG \over (3 \times 2^{15}) m_b \pi^6}
\int_{\alpha_{min}}^{\alpha_{max}} {d\alpha \over \alpha (1-\alpha)}
  \Big[m_b^2 - (1 - \alpha) \alpha s \Big]
   \Bigg\{ m_b \big[ 9 m_b^2 - 16 (1 - \alpha) m_d m_u \nnb \\
\ar 4 m_b (m_d + m_u) \big] -
    (1 - \alpha) \alpha \big[ 9 m_b + 4 (m_d + m_u) \big] s \Bigg\} \nnb \\
\ek {\gGgG \over (3 \times 2^{15}) m_b \pi^6}
\int_{\alpha_{min}}^{\alpha_{max}} {d\alpha \over \alpha^2}
 \int_{\beta_{min}}^{\beta_{max}} {d\beta \over \beta^2}
    \Big[ (\alpha + \beta) m_b^2 - \alpha \beta s\Big] \nnb \\
\cp  \Bigg\{ \Big( 58 + 5 \alpha^3 + \alpha^2 (69 - 10 \beta) -
2 \alpha \big[ 66 + \beta (23 + 22 \beta) \big] \nnb \\
\ar \beta \big[ 12 - \beta (133 + 29 \beta) \big] \Big)  m_b^3
 +   2 \beta \big[ 63 - 7 \alpha - 3 \alpha^2 - 9 (1 + \alpha) \beta -
 6 \beta^2 \big] m_b^2 (m_d + m_u) \nnb \\
\ek   2 \alpha \beta^2 (3 + 4 \alpha + 3 \beta) (m_d + m_u) s 
 + \alpha \beta m_b \Big\{ 2 (27 - 6 \alpha - 10 \beta) m_d m_u \nnb \\
\ar \big[ 24 - 3 (9 - \alpha) \alpha + 55 \beta + 7 \alpha \beta + 13 \beta^2
\big] s \Big\} \Bigg\} \nnb \\ \nnb \\
%
\rho_{12}^{(\qd)}(s) \es
  - {\qd \over 2^{10} \pi^4}
\int_{\alpha_{min}}^{\alpha_{max}} {d\alpha \over \alpha (1-\alpha)}
\Big[ m_b^2 - (1 - \alpha) \alpha s \Big] \Big[ m_b^2 (m_d - 2 m_u) \nnb \\
\ek 4 (1 - \alpha) m_b m_d m_u 
- (1 - \alpha) \alpha (m_d - 2 m_u) s \Big] \nnb \\
\ar  {\qd \over 2^{10} \pi^4}
\int_{\alpha_{min}}^{\alpha_{max}} {d\alpha \over \alpha^2}
 \int_{\beta_{min}}^{\beta_{max}} {d\beta \over \beta}
 \Big[  (\alpha + \beta) m_b^2 - \alpha \beta s \Big]
  \Bigg\{ m_b \Big( m_b \big[2 (\alpha + \beta) (1 + \alpha + \beta) m_b \nnb \\
\ar \alpha (2 + \alpha + \beta) m_d \big] -
     2 \alpha \big[ (2 + \alpha + \beta) m_b + \beta m_d \big] m_u \Big) -
   \alpha \beta \big[ 2 (1 + \alpha + \beta) m_b \nnb \\
\ar \alpha (m_d - 2 m_u) \big] s \Bigg\} \nnb \\ \nnb \\
%
\rho_{12}^{(\qu)}(s) \es
  {\qu \over 2^{10} \pi^4}
\int_{\alpha_{min}}^{\alpha_{max}} {d\alpha \over \alpha (1-\alpha)}
  \Big[ m_b^2 - (1 - \alpha) \alpha s \Big]
   \Bigg\{ m_b \Big(2 m_b m_d - \big[m_b - 4 (1 - \alpha) m_d \big] m_u \Big)\nnb \\
\ek (1 - \alpha) \alpha (2 m_d - m_u) s \Bigg\} \nnb \\
\ar {\qu \over 2^{10} \pi^4}
\int_{\alpha_{min}}^{\alpha_{max}} {d\alpha \over \alpha^2}
 \int_{\beta_{min}}^{\beta_{max}} {d\beta \over \beta}
 \Big[ (\alpha + \beta) m_b^2 - \alpha \beta s \Big]
  \Bigg\{ m_b \Big( 2 m_b \Big[ (\alpha + \beta) (1 + \alpha + \beta) m_b \nnb \\
\ek \alpha (2 + \alpha + \beta) m_d\Big] +
     \alpha \big[ (2 + \alpha + \beta) m_b - 2 \beta m_d \big] m_u \Big) -
   \alpha \beta \big[2 (1 + \alpha + \beta) m_b \nnb \\
\ek \alpha (2 m_d - m_u) \big] s\Bigg\}
\eea


\bea 
\alpha_{min} \es {s - \sqrt{ s (s - 4 m_b^2)} \over 2 s}~, \nnb \\
\alpha_{max} \es {s + \sqrt{ s (s - 4 m_b^2)} \over 2 s}~, \nnb \\
\beta_{min}  \es {m_b^2 \alpha \over s \alpha -m_b^2}~, \nnb \\
\beta_{max}  \es 1-\alpha~. \nnb
\eea


\end{document}